\documentclass{amsart}
\usepackage{amsmath}
\usepackage{amscd}
\usepackage{pb-diagram}
\usepackage{amsfonts}
\usepackage{amsthm}
\usepackage{amssymb}

\newcommand{\al}{A}
\newcommand{\circstar}{\stackrel{\star}{\circ}}
\newtheorem{thm}{Theorem}
\numberwithin{equation}{section}

\begin{document} 
\bibliographystyle{prsty}
\title{S'Darboux coordinates and WKB approximations in deformation quantization}
\author{Matthew Cargo}
\address{Department of Physics, University of California, Berkeley, California 94720 USA}
\email{mcargo@berkeley.edu}

\begin{abstract}
We introduce a method for calculating the joint spectra
of functions which comprise a quantum integrable system under
a deformation quantization star product.  The main result
involves a construction by formal power series in $\hbar$ of 
s'Darboux coordinates, a concept we introduce for
the star product analogue of Darboux normal coordinates.  
We will find that underlying the quantum integrable system is
a set of s'Darboux coordinates which in turn give rise
to number operators for the system.  We present an explicit
correction to the lowest order Bohr-Sommerfeld or EBK quantization rule.
\end{abstract}

\maketitle

\section{Introduction}

Let us describe a path from canonical quantization
to deformation quantization.
$T^*(\mathbb R)$ is a symplectic manifold with natural
coordinates $(x,p)$ and Poisson bracket
is $\{x,p\}=1$.  To quantize, we promote
$(x,p)$ to operators $(\hat x,\hat p)$ on $L^2(\mathbb R)$ with
the canonical quantization relation $[\hat x,\hat p]=i\hbar$.
One such choice has $\hat p=-i\hbar \partial/\partial x$.
We may now build 
operators out of the generators $(\hat x,\hat p)$.
Classical mechanics is left behind---only the names of 
operators $(\hat x,\hat p)$ betray their classical origin.

However, it is possible to recover the original classical
structure by using a symbol correspondence, which associates
to each operator $\hat A$ its symbol $A$, 
a function on classical phase space taking values
in power series in $\hbar$.
The operator product $\hat A\hat B$ becomes
a non-commutative product of functions $A\star B$, called a
star product.  When $A$ and $B$ are suitably well-behaved,
a star product may be developed as a power series in 
$\hbar$ of bi-differential operators.  All symbol correspondences
are such that the associated 
star commutator has $\{A,B\}_\star=
A\star B-B\star A=i\hbar \{A,B\}+O(\hbar^2)$---the 
Poisson bracket has reappeared.

Having thus returned to the classical phase space, it is natural
to ask whether the Hilbert space $L^2({\mathbb R})$ was needed
at all.  So began the so-called deformation quantization program 
of \cite{Bayen78}.  
On an arbitrary symplectic manifold, not necessary a cotangent bundle,
they sought to create a star product from scratch, assuming only that
\begin{itemize}
\item the star product is associative;
\item the star product is the ordinary product at lowest order, i.e.
$A\star B=A\cdot B+O(\hbar)$;
\item the star commutator is the Poisson bracket at lowest order, i.e.
$\{A, B\}_\star=i\hbar\{A,B\}+O(\hbar^3)$
\end{itemize}

Since the inception of deformation quantization, investigators
have affirmed the existence of a star product.
We are particularly attracted to the construction
of Fedosov~\cite{Fedosov94}, who builds a star product from the
symplectic form and a symplectic connection.

In view of these successes, one might be tempted to 
conclude that the deformation quantization program
is complete.  We disagree.
As a practical matter, many calculations that are straightforward in 
traditional Hilbert space quantum mechanics are not yet possible
in star product quantum mechanics.  In particular, we lack
systematic methods for computing spectra.
\footnote{The present form of this paper is a somewhat rushed
presentation of results which should have been published much earlier. 
I present them now to establish whatever priority they may have.
Based on a preliminary survey of the literature, I believe
the results are new. However, I am aware that many others have also pursued
work in this direction. 
In a future version intended for official publication, I fully intend
to give due credit to earlier work---hoping, of course, that
I do not find the exact result elsewhere!
}

This paper is a step in that direction. In certain types of
quantum integrable systems, we will show
that the joint spectra can be
developed as a formal power series in $\hbar$.

The organization is as follows.  In Section \ref{sec_sdarboux}, 
we will introduce
s'Darboux coordinates, the star product analogue of
Darboux normal coordinates, and show how s'Darboux coordinates
may be developed from Darboux coordinates.  In Sections \ref{sec_dirac} and
\ref{sec_qis}, we will review Dirac's creation and annihilation operator
method and define quantum integrable systems. In Section \ref{sec_good}, 
we will show that creation and annihilation operators
underlie some quantum integrable systems and develop methods
for determining these operators.  In Section \ref{sec_moyal}, we
will demonstrate these techniques with the Moyal star product
and show that they lead to higher order EBK quantization rules.
Finally, in Section \ref{sec_fedosov}, we will generalize this
result to the Fedosov star product. 

\section{Definitions and conventions}

Symbol correspondence.  Operators will wear hats, e.g. $\hat A$, and
we denote the corresponding symbol by $A=s(\hat A)$.  When
the symbol $A$ has a power series of the form $A=a+\hbar A_1+\ldots$
we call the first term, or principal
symbol, $\pi A$.  When we need to name a principal symbol, we
will try to use the corresponding lower case letter, as above where $a=\pi A$.

Lists. On our $2M$ dimensional symplectic manifold,
we shall often work with lists of $M$ or $2M$ functions.
When we refer to
a specific item in a list, we shall attach an
index, e.g. $H^i$, but when we refer to the items
collectively, we will drop the index, e.g. $H$.

Other notations will be developed as needed.

\section{From Darboux to s'Darboux coordinates}\label{sec_sdarboux}

On a $2M$ dimensional symplectic manifold ${\mathcal M}$,
Darboux normal coordinates are coordinates $z^a$ satisfying
\begin{equation}
\{z^a,z^b\}=J^{ab},
\end{equation}
where $J^{ab}$ is the {\it constant} matrix
\begin{equation}
\begin{split}
J^{ab}&=\left(
\begin{array}{cc}
0 & I_{M\times M}\\
-I_{M\times M} & 0\\
\end{array}
\right).
\end{split}
\end{equation}

A collection of $2M$ symbols $Z^i$ are said to
be s'Darboux coordinates if they have
\begin{equation}
\{Z^a,Z^b\}_\star=i\hbar J^{ab},
\end{equation}
that is, modulo the $i\hbar$, they satisfy the Darboux
coordinate algebra under the star bracket.

In this section we prove
\begin{thm}\label{thm1}  For every set of
$2M$ Darboux coordinates $z^i\in C^\infty({\mathcal M})$ there
exist s'Darboux coordinates $Z^i\in C^\infty({\mathcal M})[[\hbar]]$
such that $z^i=\pi Z^i$.  The $Z^i$ are not unique:
instead, given s'Darboux coordinates $Z^i$ there
are other sets $Z'^i=Z^i+\hbar^n X^i$ with
$\pi X^i=\{\gamma, z^i\}$ for some 
$\gamma\in C^\infty({\mathcal M})$. 
\end{thm}

The proof is by induction; we construct $Z^i$ order by order in $\hbar$.  
Already,
\begin{equation}
\{ z^i, z^j\}_\star=i\hbar J^{ij}+O(\hbar^2)
\end{equation}
since $z^i$ are Darboux coordinates.  We next assume
that  
\begin{equation}
\{Z^i,Z^j\}_\star=i\hbar J^{ij}+O(\hbar^{n+2}).
\label{Znalg}
\end{equation}
To complete the induction, we need to show there exist
$X^i\in C^\infty({\mathcal M})[[\hbar]]$ so that
\begin{equation}
\{Z^i+\hbar^{n+1}X^i,Z^j+\hbar^{n+1}X^j\}_\star=i\hbar J^{ij}+O(\hbar^{n+3}).
\label{Zn+1alg}
\end{equation}
(We then take $Z^i\rightarrow Z^i+\hbar^{n+1}X^i$.)
Rewriting the left side of Eq. \eqref{Zn+1alg}, we have
\begin{equation}
\begin{split}
\{Z^i+\hbar^{n+1}X^i,Z^j+\hbar^{n+1}X^j\}_\star&=
\{Z^i,Z^j\}_\star
+
\hbar^{n+1}\left(\{X^i,Z^j\}_\star+
\{Z^i,X^j\}_\star\right)+O(\hbar^{2n+3})\\
&=
\{Z^i,Z^j\}_\star
+
i\hbar^{n+2}\left(\{\pi X^i,Z^j\}+
\{Z^i,\pi X^j\}\right)+O(\hbar^{n+3})
\end{split}
\end{equation}
If Eq.~\eqref{Zn+1alg} is to hold, 
the terms containing $\pi X^i$ must cancel off the 
$O(\hbar^{n+2})$ part of $\{Z^i,Z^j\}_\star$.

The condition can be expressed more geometrically.
First define coordinates $z_i=J_{ij}z^j$ using the inverse $J_{ij}$ 
of $J^{ij}$.  Then define a two form,
\begin{equation}
\tilde\omega_n=\frac{1}{i\hbar}\{Z^i,Z^j\}_\star\,dz_i\wedge dz_j
\end{equation}
and a one form
\begin{equation}
\theta= X^i dz_i.
\end{equation}
Then, since 
\begin{equation}
d\theta=J^{jk}\frac{\partial X^i_{n+1}}{\partial z^k}
\, dz_j\wedge dz_i
=\{z^j, X^i\}\,dz_j\wedge dz_i,
\end{equation}
Eq.~\eqref{Zn+1alg} requires that
\begin{equation}
\tilde\omega+\frac{\hbar^{n+1}}{2}d\theta= \omega
+O(\hbar^{n+2}),
\label{geometriccondition}
\end{equation}
where $\omega=J^{ij}dz_i\wedge dz_j$ is the usual 
symplectic two form. 
Taking the exterior derivative, we obtain the condition
\begin{equation}
d\tilde\omega=\frac{1}{i\hbar}\{z^k,\{Z^i,Z^j\}_\star\} dz_k\wedge
dz_i\wedge dz_j=O(\hbar^{n+2}).
\label{2formcondition}
\end{equation}

Now, if the right side of this expression had
not a Poisson bracket but a $\star$ bracket, 
and if also $z^k$ were $Z^k$, then
it would vanish by the Jacobi identity.  
To investigate
whether these changes leave it unchanged at this order, 
we form the difference
\begin{equation}
\begin{split}
\{z^k,\{Z^i,Z^j\}_\star\}-\frac{1}{i\hbar}
\{Z^k,\{Z^i,Z^j\}_\star\}_\star
&=
\{z^k,\{Z^i,Z^j\}_\star\}-
\{Z^k,\{Z^i,Z^j\}_\star\}
+O(\hbar^{n+3})\\
&=\hbar^{n+3},
\end{split}
\end{equation}
where we have used the induction hypothesis
of Eq.~\eqref{Znalg}.
Using this calculation, we have that
\begin{equation}
\begin{split}
d\tilde \omega&=\frac{1}{i\hbar}\{z^k,\{Z^i,Z^j\}_\star\}
\, dz_k\wedge dz_i \wedge dz_j\\
&=
\frac{1}{(i\hbar)^2}
\{Z^k,\{Z^i,Z^j\}_\star\}_\star
dz_k\wedge dz_i \wedge dz_j
+O(\hbar^{n+2})\\
&=O(\hbar^{n+2}),
\end{split}
\end{equation}
where, as planned, we have used the Jacobi identity.
Now, because we have assumed that the Darboux coordinates
$z^i$ are $C^\infty$, we have also assumed that the
cohomology of ${\mathcal M}$ is trivial.  Thus, $\tilde \omega$ is
not only closed but exact at $O(\hbar^{n+2})$, and there 
exists a $\theta$ making
Eq.~\eqref{geometriccondition} true.
\footnote{One could also work on a chart on
${\mathcal M}$.  We would then have a theorem about {\it local}
s'Darboux coordinates on a manifold with arbitrary cohomology.}
Of course, $\theta$ is defined only up
to an exact one-form; adding $d\gamma$ to $\theta$ 
adds $\hbar^{n+1}\{z^i,\gamma\}$ to $Z^i$.

\section{Digression: Dirac algebra}\label{sec_dirac}

Suppose we have a set of $M$ operators $\hat A^i$ which act
on the Hilbert space $L^2({\mathbb R}^M)$ and satisfy the
algebra
\begin{equation}
\begin{split}
\left[\hat \al^{i\dagger},\hat \al^j\right]&=\hbar \delta_{ij}\\ 
\left[\hat \al^i,\hat \al^j \right] &= 0.
\end{split}
\end{equation}
The $M$ operators 
$\hat N^i=\hat \al^{i\dagger} \hat \al^i$ are
positive definite and commute.  Dirac showed the $\hat N^i$
have simultaneous eigenspaces $V_n$ labeled by
non-negative integers $n^i$: for $\psi\in V_n$,
$\hat N^i\psi=\hbar n^i\psi$.

We can attempt to duplicate this construction using
our s'Darboux symbols $Z^i$.  We create
\begin{equation}
\al^i=\frac{1}{\sqrt 2}(Z^i-i Z^{i+M})
\label{afromZ}
\end{equation}
for $i=1,\dots, M$.  These symbols satisfy Dirac's algebra under the
star bracket:
\begin{equation}
\begin{split}
\{\bar \al^i, \al^j\}_\star&=\hbar\delta_{ij}\\
\{ \al^i, \al^j\}_\star&= 0.
\label{diracalg}
\end{split}
\end{equation}
We can also define $M$ commuting symbols $N^i=\bar \al^i\star \al^i$.
\footnote{
There is no sum here.  However,
given a basis $\sigma^a$ for the Lie algebra $\mathfrak{u}(M)$, we
may define $\sigma^a_{ij}\bar \al^i\star \al^j$, a representation
of $\mathfrak{u}(M)$ under the $\star$ product.  The $N^i$ are
a maximally commuting subalgebra.
}

In order to conclude that the spectra of the $N^i$ are bounded from
below, Dirac's original argument makes use of the positive
definiteness of the $\hat N^i$.  In this case, because we do not
have a Hilbert space, 
we cannot complete Dirac's argument. 
Instead, we will
simply postulate the following: if a set of $M$
symbols $\al^i\in C^\infty({\mathcal M})[[\hbar]]$ 
satisfy Eq.~\eqref{diracalg}, then the spectrum 
of $\hbar^{-1}N^i$ consists of all non-negative integers.
\footnote{Readers are encouraged to explore if this postulate
is necessary.}

\section{Quantum integrable systems}\label{sec_qis}

In classical mechanics on a $2M$ dimensional symplectic manifold,
an integrable system is a collection of $M$ Poisson commuting, 
functionally independent functions $h^i$.
We will define a quantum integrable system as a collection
of $M$ star commuting, functionally independent functions
$H^i\in C^\infty({\mathcal M})[[\hbar]]$.  

Because $\{ H^i,H^j\}_\star=0$
implies $\{\pi H^i,\pi H^j\}=0$, we may associate to
a quantum integrable system $H^i$ the classical integrable
system $h^i=\pi H^i$.  Treatment of the quantum problem may
begin only after understanding the underlying classical mechanics.

In this paper, we will concentrate on the case where the
classical evolution takes place on tori.  To be
precise, we have the usual energy-momentum map
$h:{\mathcal M}\to {\mathbb R}^M: p\mapsto h(p)$.
For values $E$ in the interior of $h{\mathcal M}$, 
we will assume that $h^{-1}E$ is
always a single $M$-torus.  That is,
the $h^i$ give rise to a global set of
action-angle variables $(\theta^1,\ldots,\theta^M,I^1,\ldots,I^M)$
satisfying
\footnote{
In $I^i$, we have an exception to using the
lower case for classical functions.
}
\begin{equation}
\begin{split}
\{\theta^i,\theta^j\}&=0\\
\{I^i,I^j\}&=0\\
\{\theta^i,I^j\}&=\delta^{ij},
\end{split}
\label{aaalg}
\end{equation}
and functions $f^i:{\mathbb R}^M\rightarrow {\mathbb R}$
which express the $h^i$ in terms of the $I^i$:
\begin{equation}
h^i=f^i\circ I.
\end{equation}

In view of Eq.~\eqref{aaalg}, it is tempting to 
regard the $(\theta,I)$ as Darboux coordinates.  This
we should not do, however, as the $\theta^i$ are not
$C^\infty$.  Nevertheless, we may create Darboux coordinates
from this set by defining
\begin{equation}
\begin{split}
z^i&=\sqrt{2I^i}\cos(\theta^i)\\
z^{i+M}&=-\sqrt{2I^i}\sin(\theta^i);
\end{split}
\end{equation}
it is straightforward to verify $\{z^i,z^j\}=J^{ij}$.  In
doing so we must assume that the basis contours used to define
the $I^i$ have been chosen so that $I^i\ge 0$ everywhere.
In other work, my collaborators will argue that such contours
do exist for this type of integrable system. 
In addition, we shall assume
the $\theta^i$ can be and indeed are chosen so that the
$z^i$ are $C^\infty$.\footnote{
The angle coordinates $\theta^i$ are, of course, not
unique.  Taking $\theta^i\rightarrow \theta^i + 
\{\theta^i, \gamma\}$ for some function $\gamma$ gives
new angle coordinates.}

Previously, we proved Darboux coordinates $z^i$
can be quantized
to s'Darboux symbols $Z^i$ satisfying $z^i=\pi Z^i$.  In
Eq.~\eqref{afromZ} and following, we showed how to create number operators
$N^i$ from the $Z^i$.
Next, we will show that the
$Z^i$ can be made so that $\{H^i,N^j\}_\star=0$; when
this condition holds, we shall say the $N^i$ are good number operators
for $H^i$ and that the $Z^i$ are compatible with $H^i$.
\footnote{Already, it is possible to {\it construct}
quantum integrable systems from a classical one:
from $h^i$ find $(I,\theta)$, form $z^i$, quantize $z^i$ to $Z^i$,
form number symbols $N^i$, and then define $H^i=h^i\circstar N$.
Because the $Z^i$ are not unique, neither are the resulting $H^i$.}

\section{Existence of good number operators}\label{sec_good}

In this section, we prove two theorems.  The first  
concerns ``approximately
good'' number operators.  It will help prove the second, our main result.

We first introduce some new notation. Let $f$ be an analytic function
with Taylor series $f(x)=\sum c_n x^n$.  Define
$f\circstar A=\sum c_n A^{\star n}$; that is, for each term
$x^n$, we substitute the $n$-fold star product of $A$.  If $A$ is the image
of $\hat A$ under a symbol correspondence, then $f\circstar A=
s(f(\hat A))$.  Thus the $\circstar$ notation gives us a way to talk
about the symbol of a function of an operator in the context of 
deformation quantization, where there is neither a Hilbert space 
nor operators on it.

For the purposes of this section, the only fact we need to know
about $\circstar$ is that for $\star$ commuting functions  
$N^i$, $F\circstar N=F\circ N+O(\hbar)$.  This fact follows immediately from
the deformation quantization postulates.

The two theorems can now be stated as
\begin{thm}\label{thm2}
Suppose $H^i$ and $N^i$ 
satisfy $\{H^i,H^j\}_\star=0$,
$\{N^i,N^j\}_\star=0$, $\{N^i,H^j\}_\star=O(\hbar^{n+2})$, and
there are $f^i$ such that $\pi H^i=f^i\circ\pi N$.
Then
there exist $F^i\in C^\infty({\mathbb R}^M)[[\hbar]]$ and
$G^i\in C^\infty({\mathcal M})[[\hbar]]$ such that
$H^i=F^i\circstar  N+\hbar^{n+1}G^i.$ 
Furthermore, $\pi G^i=\{G,h^i\}+g^i\circ I$, for
some $G,g^i\in C^\infty({\mathcal M})$.
\end{thm}

\begin{thm}\label{thm3}
There exist s'Darboux symbols $Z^i$ compatible with
the quantum integrable system $H^i$.
\end{thm}

Before proving these we remark that
nothing in Theorem \ref{thm2} requires that the
$N^i$ are number symbols.  However, if they are,
the spectrum of ~$H^i=F^i\circstar N+O(\hbar^n)$
is just $F^i\circ(\hbar n)+O(\hbar^n)$, where $n_i$ 
are non-negative integers.

\subsection{Proof of Theorem \ref{thm2}}
We prove the theorem by induction, constructing 
$F^i$ order by order.  We begin with
$F^i=f^i$.  Then 
\begin{equation}
F^i\circstar N=f^i\circ N+O(\hbar)
=f^i\circ \pi N+O(\hbar)=H^i+O(\hbar).
\end{equation}
We next assume that $F^i$ is such that
\begin{equation}
F^i\circstar N=H^i+O(\hbar^{s+1}).
\label{sfi}
\end{equation}
To complete the induction, we need to show there exist
$X^i\in C^\infty({\mathbb R}^M)$ such that
\begin{equation}
(F^i+\hbar^{s+1}X^i)\circstar N=H^i+O(\hbar^{s+2})
\label{sfi2}
\end{equation}
whenever $s\le n-1$.  (We then take $F^i\rightarrow F^i+\hbar^{s+1}X^i$.)
Working with the left side of Eq.~\eqref{sfi2}, the condition becomes
\begin{equation}
F^i\circstar N+\hbar^{s+1}X^i\circ \pi N=H^i+O(\hbar^{s+2}).
\end{equation}

Because $X^i\circ\pi N$ depends only on the $N^i$,
$X^i$ may be used to enforce this equation only when the
other terms also depend only on the $N^i$.
To verify this, we compute
\begin{equation}
\begin{split}
\{\pi N^j,F^i\circstar N-H^i\}
&=\frac{1}{i\hbar}\{N^j,F^i\circstar N-H^i \}_\star+O(\hbar^{s+2})\\
&=
\frac{1}{i\hbar}\left(0+O(\hbar^{n+2})\right)+O(\hbar^{s+2})\\
&=
O(\hbar^{n+1})+O(\hbar^{s+2}).
\end{split}
\end{equation} 
This is of $O(\hbar^{s+2})$ whenever $s\le n-1$.
At the final stage of the induction process, we have $s=n-1$ and
\begin{equation}
H^i=F^i\circstar N+\hbar^{n+1}G^i,
\label{decomp}
\end{equation}  
for some $G^i$.  

This concludes the first part of theorem \ref{thm2}.
For the second part, we compute
\begin{equation}
\begin{split}
0&=\{H^i,H^j\}_\star\\
&=\{F^i\circstar N+\hbar^{n+1}G^i,F^j\circstar N+\hbar^{n+1}G^j\}_\star\\
&=\hbar^{n+1}(\{F^i\circstar N,G^j\}_\star
+\{G^i,F^j\circstar N\}_\star)+O(\hbar^{2n+2})\\
&=i \hbar^{n+2}(\{f^i\circ\pi N,\pi G^j\}
+\{\pi G^i,f^j\circ \pi N\})+O(\hbar^{n+3}).
\end{split}
\end{equation}
Thus, since $\{\pi H^i,\pi G^j\} +\{\pi G^i,\pi H^j\}=0$, we have
$\pi G^j=\{G,\pi H^j\}+g^j\circ \pi I$ for some functions
$G$ and $g^j$.  Note that
$G$ is not determined uniquely in this decomposition---we
could add a function of the $\pi N^i$. With this result, 
equation (\ref{decomp})
becomes
\begin{equation}
H^i=F^i\circstar N+\hbar^{n+1}(\{G,\pi H^i\}+g^j\circ\pi N)+
O(\hbar^{n+2}).
\label{decomp2}
\end{equation}
This completes theorem \ref{thm2}.

\subsection{Proof of Theorem \ref{thm3}}

We will construct the symbols $Z^i$ order by order
in $\hbar$, taking care that the associated
number symbols $N^i$ commute with the $H^i$ at
the current order.
We assume that $Z^i$ are s'Darboux symbols, but that
$\{\bar \al^i\star \al^i, H^j\}_\star=O(\hbar^{n+2}).$
For the induction step, we will create new d'Darboux
symbols
\begin{equation} 
Z'^{i}=Z^i+\hbar^{n+1}g^i,
\label{Zprime}
\end{equation} 
so that the associated $N'^i=\bar \al'^i\star \al'^i$ satisfy 
$\{\bar \al'^{i}\star \al'^{i}, H^j\}_\star=O(\hbar^{n+3}).$

As discussed before, the
the freedom in the $Z^i$, allows
us to take $\pi g^i=\{\gamma,z^i\}$,
with $\gamma$ a real function.  This
modifies the number operators
at order $\hbar^{n+1}$.  First,
\begin{equation}
\al'^i=\frac{1}{\sqrt 2}\left(Z'^i-iZ'^{i+M}\right)=
\al^i+\hbar^{n+1}\{\gamma,\pi \al^i\}+O(\hbar^{n+2}).
\end{equation}
Then
\begin{equation}
\begin{split}
N'^{i}&=
\left(\bar \al^i+\hbar^{n+1}\{\gamma,\pi \bar \al^i\}+O(\hbar^{n+2})\right)
\star 
\left(\al^i +\hbar^{n+1}\{\gamma,\pi \bar \al^i\}+O(\hbar^{n+2})\right)\\
&= N^i+
\hbar^{n+1}\left(\pi \bar\al^i\{\gamma,\pi \al^i\}+\pi \al^i\{\gamma,\pi\bar \al^i\}\right)
+O(\hbar^{n+2})\\
&= N^i+\hbar^{n+1}\{\gamma, I^i\}+O(\hbar^{n+2})
\label{Nprimeeqn}
\end{split}
\end{equation}
where we have used that $(\pi \bar \al^i)(\pi \al^i)=I^i$.

We want to choose $\gamma$ so that
$\{N'^{i},H^j\}_\star=O(\hbar^{n+3})$.  Computing, we have
\begin{equation}
\begin{split}
\{N'^{i},H^j\}_\star&=
\{N^i+\hbar^{n+1}\{\gamma, I^i\}\,,
s(F^j\circ\hat N)+\hbar^{n+1}(\{G,h^j\}+g^j\circ I)
)\}_\star+O(\hbar^{n+3})\\
&=i\hbar^{n+2}\{\{\gamma, I^i\},h^j\}+
i\hbar^{n+2}\{I^i,\{G,h^j\}+g^j\circ I\}+O(\hbar^{n+3})\\
&=i\hbar^{n+2}\left(
\{\{\gamma, I^i\},h^j\}+
i\hbar^{n+2}\{I^i,\{G,h^j\}\right)+O(\hbar^{n+3}).
\end{split}
\end{equation}
where we have used the decomposition of theorem \ref{thm2}.
Choosing $\gamma=G$ and using the Jacobi identity,
we finish theorem \ref{thm3}.

Remark. $\gamma$ is determined only up to a function of the
actions, and thus the $\al^i$ are not determined uniquely. 
To what does this freedom correspond on the quantum level?
Consider the following
transformation of the $\hat \al^i$:
\begin{equation}
\hat \al^i\to e^{-i\hat G\circ \hat N/\hbar}\hat \al^i 
e^{i\hat G\circ \hat N/\hbar},
\label{quantangle}
\end{equation}
where $G\in C^\infty({\mathbb R}^M)[[\hbar]]$.
This transformation leaves the $\hat N^i$ and the $\hat A$ algebra
invariant. On the level of wavefunctions, this transformation
of the creation operators amounts to an $\hat N^i$ dependent rephasing.

Using the commutation relations, the transformation can also be written
\begin{equation}
\hat \al^i\to \hat \al^i 
e^{i\hat \hat G_i\circ \hat N/\hbar},
\label{quantangle2}
\end{equation}
with $\hat G_i=G\circ(\hat N+\hbar/2)-G\circ(\hat N+\hbar/2-\hbar e_i)$.
At lowest order in $\hbar$, we have $s(\hat G_i)/\hbar=(\partial_i G)\circ I=
\{\theta^i,G\circ I\},$ and thus
$\al^i\to \al^i e^{i\{\theta^i,\,G\circ I\}}.$  This
is just the classical freedom in the origin of the angle variables.
The indeterminacy of $\gamma$ is
the manifestation in Eq.~(\ref{quantangle}) at 
higher order in $\hbar$.  

Of course, Eq.~(\ref{Nprimeeqn}) shows the $N^i$ symbols
themselves are uniquely determined at each order.  Later, 
when we work with the Moyal star product, we will see 
directly that the $O(\hbar^2)$ part of $N^i$ 
is uniquely determined from the $H^i$.

\section{Example: Quantization of Moyal Symbols}\label{sec_moyal}

In this section, we will specialize to the Moyal star
product\cite{Moyal} and show how our prior analysis leads to
higher order Bohr-Sommerfeld quantization rules for
some quantum integrable systems of Weyl symbols\cite{Weyl}.

The assumptions are now as follows: we have a
set of $M$ symbols $H^i$ which Moyal commute and whose
principal symbols $h^i=\pi H^i$ give rise to a 
global set of action angle variables $(I,\theta)$.  We
suppose that the series for $H^i$ are even in $\hbar.$  Further,
we suppose that the functions $z^i=\sqrt{2I^i}\cos{\theta^i}$
and $z^{i+M}=-\sqrt{2I^i}\sin{\theta^i}$ are $C^\infty$ and are
Darboux coordinates.

To achieve a quantization rule beyond the
usual WKB order, we need to

\begin{itemize}
\item  extend the $z^i$ to symbols $Z^i$ satisfying
$\pi Z^i=z^i$ and $\{Z^i, Z^j\}_\star=i\hbar J^{ij}+O(\hbar^5)$.

\item modify the $Z^i$ so that the associated
number operators $N^i$ satisfy $\{H^i,N^j\}_\star=O(\hbar^5)$.
\end{itemize}

This is the work of section \ref{secmoyal1}
and \ref{secmoyal2}, respectively.
The first problem will require some techniques.  Before
developing them, we must review the
Moyal star product.

\subsection{The Moyal star product}

The Moyal star product is a star product on
${\text T}^*({\mathbb R}^{M})$ phase space, with the standard
Poisson bracket 
\begin{equation}
\{f,g\}=J^{ij}\partial_i f\partial_j g=\partial_i f\partial^i g.
\footnote{
We use $J$ to raise and lower indices.  For example
$T^i=J^{ij}T_j$ and $T_i=J_{ij}T^j$.
}
\end{equation}
Here, the partial derivatives $\partial_i$ are with respect to the
standard $(x,p)$ coordinates on $T^*({\mathbb R}^{M})$,
and $J$ is as before.
The star product, for $f,g\in C^\infty({\mathbb R}^{2M})[[\hbar]]$,
is given by the Moyal formula 
\begin{equation}
f\star g=\sum_{n=0}^\infty
\frac{1}{N!}
\left(\frac{i\hbar}{2}\right)^n
\{f,g\}_n
\end{equation}
where 
\begin{equation}
\begin{split}
\{f,g\}_0&=f\cdot g\\
\{f,g\}_1&=\{f, g\}=\partial_i f \partial^i g\\
\{f,g\}_2&=\partial_{i_1}\partial_{i_2} f 
\partial^{i_1}\partial^{i_2}g
\label{ugly}
\end{split}
\end{equation}
and so on.   The Moyal commutator $\{f,g\}_\star=f\star g-g\star f$
is an odd series in $\hbar$ and because of this, as we shall see,
we can develop expansions for $Z^i$, $N^i$, etc as even series. 
Accordingly, the results from Sections \ref{sec_sdarboux} and \ref{sec_good},
will differ somewhat.

The notation of Eq.~(\ref{ugly}) is not always convenient.
To avoid proliferating indices, 
we will resume 
a diagrammatic notation introduced earlier in \cite{moyalbs1}.
We define $f\rightarrow g=\{f, g\}$, and, for more complicated 
diagrams, we exhibit an example conversion of an arrow to
more explicit notation:
\begin{equation}
\begin{split}
\begin{diagram}
\node[2]{D}
\arrow{s}\\
\node{A}
\arrow{e}
\node{B}
\arrow{e}
\node{C}
\end{diagram}&=
\begin{diagram}
\node{\partial_i D \, A
}
\arrow{e}
\node{\partial^i B}
\arrow{e}
\node{C}
\end{diagram}
\end{split}
\end{equation}

\subsection{Construction of Moyal s'Darboux Coordinates}\label{secmoyal1}

We need symbols $Z^i=z^i+\hbar^2 Z_2^i+\ldots$
satisfying 
\begin{equation}\label{2ndordereqn}
\{Z^i,Z^j\}_\star=i\hbar J^{ij}+O(\hbar^5).
\end{equation}
We have found such $Z^i_2$.
First define
\footnote{The fact that the $\Gamma^{abc}$ 
are the connection coefficients for the flat symplectic
connection in $z^i$ coordinates is not central to
the derivation, but it is highly suggestive, especially
in view of Fedosov's employ \cite{Fedosov94} of a symplectic connection
in constructing his  star product. 
For now, we simply note that $\Gamma^{abc}$ is 
completely symmetric in its three indices.}
\begin{equation}
\begin{split}
\Gamma^{abc}&=
\begin{diagram}
\node{z^a}
\arrow{e}
\node{z^b}
\node{z^c}
\arrow{w}
\end{diagram}.
\end{split}
\end{equation}
Then
\begin{equation}
Z^i_2 =\frac{1}{48}\Gamma_{abc}\, z^i\rightarrow \Gamma^{abc}
\label{Z2eqn}
\end{equation}
satisfies Eq.~(\ref{2ndordereqn}).
\footnote{Despite the known non-uniqueness of the $Z_2$, one wonders
if this construction is canonical in some sense.
}

The key to verifying this will be the following
identity, which we present without motivation:
\begin{equation}
\begin{split}
\{z^d,\Gamma^{abc}\}
&=
\begin{diagram}
\node[2]{z^b}
\arrow{s}
\\
\node{z^a}
\arrow{e}
\node{z^d}
\node{z^c}
\arrow{w}
\end{diagram}
\label{magicident}.
\end{split}
\end{equation}
The proof is as follows. 
Because
the bracket $\{z^b,z^d\}$ is constant, we have
\begin{equation}
\begin{split}
0&=
\begin{diagram}
\node{z^a}
\arrow{e}
\node{(z^b}
\arrow{e}
\node{z^d)}
\arrow{e}
\node{z^c}
\end{diagram}\\
&=
\begin{diagram}
\node{z^a}
\arrow{e}
\node{z^b}
\arrow{e} 
\node{z^d}
\arrow{e}
\node{z^c}
\end{diagram}
+
\begin{diagram}
\node{z^a}
\arrow{e}
\node{z^d}
\node{z^b}
\arrow{w}
\arrow{e}
\node{z^c}
\end{diagram}\\
&
+
\begin{diagram}
\node[2]{z^d}\\
\node{z^a}
\arrow{e}
\node{z^b}
\arrow{n}
\arrow{e}
\node{z^c}
\end{diagram}
+ 
\begin{diagram}
\node[2]{z^b}
\arrow{s}\\
\node{z^a}
\arrow{e} 
\node{z^d}
\arrow{e}
\node{z^c}
\end{diagram}
\nonumber\\
&=
\begin{diagram}
\node{z^d}
\node{(z^a}
\arrow{e}
\arrow{w}
\node{z^b}
\arrow{e}
\node{z^c)}
\end{diagram}
+
\begin{diagram}
\node[2]{z^b}
\arrow{s}\\
\node{z^a}
\arrow{e}
\node{z^d}
\arrow{e}
\node{z^c}
\end{diagram}.
\end{split}
\end{equation}
With rearrangement, the identity follows.

A resolution
of the arrow,
\begin{equation}
\to\,\,=\,\,\to z^a\, z_a\to,
\end{equation}
will also prove useful.
At order $\hbar^3$ in $\{Z^i,Z^j\}_\star$ we have
\begin{equation}
i\hbar^3
\left(\left(
\{z^i, Z^j_2\}-i\leftrightarrow j\right)
-\frac{1}{24}
\{z^i,z^j\}_3.
\right)
\end{equation}
According to Eq.~(\ref{2ndordereqn}), this must vanish.  Working with the $Z^i_2$ terms, we have
\begin{equation}
\begin{split}
\{z^i, Z^j_2\}-i\leftrightarrow j
&=
\frac{1}{48}
\{z^i,\Gamma_{abc}\{z^j, \Gamma^{abc}\}\}-i\leftrightarrow j\\
&=
\frac{1}{48}\left(
\{z^i,\Gamma_{abc}\}\{z^j, \Gamma^{abc}\}+
\Gamma_{abc}\{z^i,\{z^j, \Gamma^{abc}\}\}\right)
-i\leftrightarrow j\\
&=
\frac{1}{48}
\{z^i,\Gamma_{abc}\}\{z^j, \Gamma^{abc}\}
-i\leftrightarrow j\\
&=
\frac{1}{24}
\{z^i,\Gamma_{abc}\}\{z^j, \Gamma^{abc}\}\\
&=
\frac{1}{24}
\begin{diagram}
\node[2]{z_b}
\arrow{s}\\
\node{z_a}
\arrow{e}
\node{z^i}
\node{z_c}
\arrow{w}
\end{diagram}
\begin{diagram}
\node[2]{z^b}
\arrow{s}\\
\node{z^a}
\arrow{e}
\node{z^j}
\node{z^c}
\arrow{w}
\end{diagram}\\
&=
\frac{1}{24}
\{z^i,z^j\}_3
\end{split}
\end{equation}

We have used the Jacobi identity, the identity~(\ref{magicident}),
and the resolution of the arrow.  Equation~(\ref{2ndordereqn})
is verified.

We now have an explicit form for the
first two terms in $Z^i=z^i+\hbar^2 Z^i_2+\ldots$.  According
to Theorem \ref{thm1}, the remaining terms in the
$Z^i$ series can be completed so that the $Z^i$ are s'Darboux coordinates.
To make the forthcoming analysis clearer, we shall now assume 
this has been done, although none of the results 
depend on the specific form of $Z^i_4$, etc.

\subsection{Good number operators}\label{secmoyal2}

Associated to our $Z^i$ are number operators $N^i$.
They are easily expressed in terms of $\al^i=a^i+\hbar^2 \al^i_2+O(\hbar^4)$.
\begin{equation}
\begin{split}
N^i&=\bar \al^i\star \al^i\\
&= \bar a^i a^i + \frac{i\hbar}{2}\{\bar a^i,a^i \}
-\frac{\hbar^2}{8}\{\bar a^i,a^i\}_2
+\hbar^2(\bar \al^i_2 a^i+\bar a^i \al^i_2)
+O(\hbar^3)\\
&=I^i-\frac{\hbar}{2}
-\frac{\hbar^2}{8}\{\bar a^i,a^i\}_2
-\frac{\hbar^2}{48}\Gamma_{abc}I^i\to\Gamma^{abc}
+O(\hbar^3)
\end{split}
\end{equation}
The last two terms define 
\begin{equation}
N^i_2=-\frac{1}{8}\{\bar a^i,a^i\}_2
-\frac{1}{48}\Gamma_{abc}I^i\to\Gamma^{abc}.
\end{equation}
The appearance of an $O(\hbar)$ term
\footnote{
This term is, of course, related to the Maslov index.
}
means that will be more convenient to develop
$F^i$ as
$H^i=F^i\circstar(N^i+\hbar/2)$.

Already, we have $\{N^i,H^j\}_\star=O(\hbar^3)$.  
We would like to modify $Z^i$ so that
$\{N^i,H^j\}_\star=O(\hbar^5)$.  
Taking $F^i=f^i$, 
\footnote{Recall that $f^i$ is $h^i$ expressed as 
a function of its action variables, i.e. 
$f^i\circ I=h^i$.}
we have $H^i-F^i\circstar (N+\hbar/2)=O(\hbar^2)$. 
By the second part of Theorem \ref{thm2}, 
$H^i=f^i\circstar ( N+\hbar/2)+\hbar^2 G^i$ for $G^i$ 
in which $\pi G^i=\{G,h^i\}+g^i\circ I$.
The $\{G,h^i\}$ term in $\pi G^i$ may be isolated subtracting
the angle independent terms from 
$H^i-f^i\circstar( N+\hbar/2)$.  Defining a complement
to the angle average by $\left>f\right<=f-\left<f\right>$,
we have
$\{G,h^i\}=\left>H^i-f^i\circstar ( N+\hbar/2)\right<$.

Using a more accurate expansion of $\circstar$, 
\cite{CdV}
we have
\begin{equation}
\begin{split}
f^i\circstar ( N+\hbar/2)&=
h^i-
\frac{\hbar^2}{16} 
\{I^j,I^k\}_2 
(\partial_j \partial_k f^i) \circ I\\
&-\frac{\hbar^2}{24} 
(I^j\to I^k\gets I^l)
(\partial_j \partial _k\partial_l f^i)\circ I
 +\hbar^2 N^j_2(\partial_j f^i)\circ I+O(\hbar^4)\\
&= h^i+\hbar^2 K^i_2+\hbar^2\Omega_{ij} N^j_2 +O(\hbar^4),
\end{split}
\end{equation}
where, for convenience, we have defined 
the frequency matrix $\Omega_{ij}=(\partial_j f^i)\circ I$ 
and
\begin{equation}
K^i_2=\frac{\hbar^2}{16} 
\{I^j,I^k\}_2 
(\partial_j \partial_k f^i) \circ I+
\frac{\hbar^2}{24} 
I^j\to I^k\gets I^l\,
(\partial_j \partial _k\partial_l f^i)\circ I.
\end{equation}

We now have
\begin{equation}
\{G, h^i\}=\frac{\partial G}{\partial\theta^k}\Omega_{ik}
=
\left>
H^i_2- K^i_2-N^j_2 \Omega_{ij}\right<,
\end{equation}
and may solve for $\partial G/\partial \theta^k$:
\begin{equation}
\frac{\partial G}{\partial \theta^i}=
\left>
\Omega^{-1}_{ik}(H^k_2- K^k_2)\right<-\left>N^i_2\right<.
\end{equation}

As in Theorem \ref{thm3}, we define 
$Z'^i=Z^i+\hbar^2\{\gamma, z^i\}+O(\hbar^4)$,
\footnote{
Here we have one extra power of $\hbar$.
}
and choose $\gamma=G$, 
so that $\{\bar \al'^i\star \al'^i,H^j\}_\star=O(\hbar^5)$.
The new number operators are given by Eq.~(\ref{Nprimeeqn}):
\begin{equation}
\begin{split}
N'^i+\hbar/2&=N^i+\hbar/2+\hbar^2\{G,I^i\} + O(\hbar^4)\\
&=I^i+\hbar^2 N^i_2+\hbar^2\frac{\partial G}{\partial \theta^i}\\
&=I^i+\hbar^2 N^i_2+\hbar^2
\left>\Omega^{-1}_{ik}(H^k_2- K^k_2)\right<-\hbar^2\left>N^i_2\right<\\
&=I^i+\hbar^2\left< N^i_2\right>+\hbar^2
\left>\Omega^{-1}_{ik}(H^k_2- K^k_2)\right<
\end{split}
\end{equation}
Note that $\langle N'^i_2\rangle=\langle N^i_2\rangle.$

Finally, we can obtain the quantization condition.  
We require
\begin{equation}
\begin{split}
H^i&=h^i+\hbar^2 H^i_2+O(\hbar^4)=F^i\circstar( N'+\hbar/2)+O(\hbar^4)\\
&=f^i\circstar(N'+\hbar/2)+\hbar^2 F^i_2\circstar( N'+\hbar/2)+O(\hbar^4)\\
&=h^i+\hbar^2 K^i_2+\hbar^2\Omega_{ij}N'^j_2+\hbar^2
+\hbar^2 F^i_2\circ I+O(\hbar^4)
\end{split}
\end{equation}
By the work of Theorem \ref{thm2}, we know this equation is satisfied by taking
\begin{equation}
\begin{split}
F^i_2\circ I&=\langle H^i_2-K^i_2-\Omega_{ij}N'^j_2\rangle\\
&=\langle H^i_2-K^i_2-\Omega_{ij}N^j_2\rangle.
\end{split}
\label{bsquant}
\end{equation}

$F=f^i+\hbar^2 F^i_2$ is the second order Bohr-Sommerfeld quantization
rule for quantum integrable systems of Moyal symbols.  
Equation~(\ref{bsquant}) contains two important terms: the
$K^i_2$ comes from knowing the symbol of a function of
an operator, while the $N^i_2$ term comes directly from
the s'Darboux construction.

\section{Construction of s'Darboux coordinates in the Fedosov star product}\label{sec_fedosov}

Let ${\mathcal M}$ be a manifold with symplectic 
form $\omega$ and a torsionless symplectic connection
$\nabla$, i.e. one with $\nabla\omega=0$.  
From $(\omega,\nabla)$, Fedosov
constructs a star product.\cite{Fedosov94} 
To summarize his result, given any 
$f\in C^\infty ({\mathcal M})$ and Darboux local coordinates $z^i$,
there are functions $f^{(n)}_{i_1,\ldots,i_n}$ which
are completely symmetric in the numbered indices
\footnote{For this section, we establish a convention 
that implicit any expression
containing numbered indices is a symmetrization over those indices.}
such that Fedosov's star product is
\begin{equation}
f\star g =
f^{(0)}g^{(0)}+
\frac{i\hbar}{2}
f^{(1)}_{i_1}\omega^{i_1j_1}g^{(1)}_{j_1}+\cdots+
\frac{1}{n!}\left(\frac{i\hbar}{2}\right)^n
f^{(n)}_{i_1,\ldots,i_n}\omega^{i_1j_1}\cdots\omega^{i_n j_n}
g^{(n)}_{j_1,\ldots,j_n}+\ldots
\end{equation}
where $\omega^{ij}$ is the Poisson tensor with respect
to the $z^i$.  For early $f^{(n)}$, the results are
\begin{equation}
\begin{split}
f^{(0)}&=f\\
f^{(1)}_{i_1}&=\nabla_{i_1} f\\
f^{(2)}_{i_1,i_2}&=\nabla_{i_1}\nabla_{i_2} f\\
f^{(3)}_{i_1,i_2,i_3}&=\nabla_{i_1}\nabla_{i_2}\nabla_{i_3} f-
\left. R_{i_1i_2i_3}\right.^j \nabla_j f,
\end{split}
\end{equation}
where the derivatives are taken with respect
to the $x^i$ coordinates, and 
where $\left. R_{i_1 i_2 i_3}\right.^j=
\omega_{i_1 i}\left. R^i\right._{i_2 i_3 k}\omega^{jk}$ is
the Riemann curvature tensor in those coordinates.~\footnote{
It seems Fedosov's own expression for $f^{(3)}$ contains
an error. He has $1/4$ as the coefficient of $R$; we think
it is just $1$.}

Given Darboux coordinates $z^a$,
we wish to construct s'Darboux symbols $Z^a=z^a+\hbar^2 Z^a_2$.
It is natural to guess that $Z^a_2$ is as in 
Eq.~(\ref{Z2eqn}), but with $\Gamma$ the new, possibly
curved, symplectic connection written in the $z^a$ coordinates.
We shall verify this.

First, we need to prove a generalization
of Eq.~(\ref{magicident}):
\begin{equation}
\nabla_{a_1} \nabla_{a_2} \nabla_{a_3} z^d -
\left. R_{a_1 a_2 a_3}\right.^d
=-\partial^d\Gamma_{a_1 a_2 a_3}
\label{gmagicident}
\end{equation}
where $\Gamma^{abc}$ are the coefficients of the symplectic connection 
in $z^a$ coordinates. 

To prove this, we need only the fact that when
$\Gamma^{abc}$ for a symplectic connection is written
in Darboux coordinates, it is completely symmetric 
in its indices.\cite{Fedosov94} 
Let $\partial_{a}=\partial/\partial z^a$.  First, we have
\begin{equation}
\begin{split}
\nabla_{a_1} \nabla_{a_2} \nabla_{a_3} z^d
&=
\nabla_{a_1} \nabla_{a_2} \delta^d_{a_3}\\
&=
-\nabla_{a_1}\Gamma^d_{a_3a_2}\\
&=
-\partial_{a_1}\Gamma^d_{a_3a_2}+
\Gamma^\sigma_{a_1 a_3}\Gamma^d_{\sigma a_2}+
\Gamma^\sigma_{a_1 a_2}\Gamma^d_{\sigma a_3}\\
&=
-\partial_{a_1}\Gamma^d_{a_3a_2}+
2\Gamma^\sigma_{a_1 a_3}\Gamma^d_{\sigma a_2},
\end{split}
\end{equation}
where, in the last step, we have used the implicit
symmetrization on numbered indices and the symmetry
of $\Gamma^{abc}$.  Using the same property of $\Gamma^{abc}$,
we also have
\begin{equation}
\begin{split}
\left. R_{a_1 a_2 a_3}\right.^d&=-
\partial_{a_3}\Gamma^d_{a_1 a_2}+
\partial^d\Gamma_{a_1 a_2 a_3}
+
\Gamma^\eta_{a_1 a_3}\Gamma^d_{\eta a_2}
+\Gamma^\eta_{a_2 a_3}\Gamma^d_{\eta a_1}\\
&=-\partial_{a_3}\Gamma^d_{a_1 a_2}+
\partial^d\Gamma_{a_1 a_2 a_3}
+2 \Gamma^\eta_{a_1 a_3}\Gamma^d_{\eta a_2}.
\end{split}
\end{equation}

Subtracting these two results, we obtain Eq.~(\ref{gmagicident}).
By a calculation very similar to the one in section \ref{secmoyal1},
we can show that
\begin{equation}
Z^d=z^d+\frac{\hbar^2}{48}\Gamma_{abc}\{z^d,\Gamma^{abc}\}
\label{fedosovstarboux}
\end{equation}
satisfy $\{Z^i,Z^j\}=i\hbar J^{ij}+O(\hbar^5)$.

Following our previous development, we obtain the EBK formula in the 
Fedosov quantization by replacing
the $\Gamma$ in Eq. \eqref{bsquant} with this new $\Gamma$, and
by changing the ordinary derivatives 
to covariant derivatives.

\section{Conclusion}
In this paper, we have developed a 
method for calculating spectra of functions which comprise a quantum
integrable system under a star product.

The methods developed here may have other applications.
A future, more official, presentation may also
include thoughts about
\begin{itemize}
\item star product quantization rules on cohomologically non-trivial manifolds;
\item star Lie algebras from Poisson Lie algebras and applications
to the semi-classics of spin;
\item Heisenberg evolution of s'Darboux coordinates.
\end{itemize}

\section{Acknowledgments}
Robert Littlejohn and Alfonso
Gracia-Saz have provided valuable insight and feedback throughout the 
development of this work.

\bibliography{ReducedStarboux}

\end{document}